\documentclass[journal,10pt]{IEEEtran}

\usepackage[nolist]{acronym}
\begin{acronym}
  \acro{RIS}{reconfigurable intelligent surface}
  \acro{FRIS}{fluid reconfigurable intelligent surface}
  \acro{FRISs}{fluid reconfigurable intelligent surfaces}
  \acro{FAS}{fluid antenna system}
  \acro{SNR}{signal-to-noise ratio}
  \acro{BS}{base station}
  \acro{UE}{user equipment}
  \acro{UPA}{uniform planar array}
  \acro{PDF}{probability density function}
  \acro{CDF}{cumulative distribution function}
  \acro{LOS}{line-of-sight}
  \acro{Tx}{transmitter}
  \acro{Rx}{receiver}
  
\end{acronym}

\usepackage{amssymb,amsmath,amsfonts,amsthm,bm,bbm}
\usepackage{empheq}
\usepackage{algorithm}
\usepackage[noend]{algpseudocode} 

\usepackage[caption=false,font=normalsize,labelfont=sf,textfont=sf]{subfig}
\usepackage{graphicx}
\usepackage{mathrsfs}
\usepackage{xcolor}

\usepackage{array}
\usepackage{stfloats}
\usepackage{url}
\usepackage{verbatim}

\usepackage{cite}
\usepackage[citecolor=black]{hyperref}

\usepackage{acronym}

\begin{document}

\title{Performance Limits of FRIS Systems in Nakagami-$m$ Fading}

\author{Masoud Khazaee, Felipe A. P. de Figueiredo, Rausley A. A. de Souza, \IEEEmembership{Senior Member,~IEEE}, \\ Farshad Rostami Ghadi, \IEEEmembership{Senior Member,~IEEE}, Kai-Kit Wong, \IEEEmembership{Fellow,~IEEE}, Luciano L. Mendes,~%
\IEEEmembership{Member,~IEEE}, \\ and Fernando D. Almeida García,~\IEEEmembership{Senior Member,~IEEE}

\thanks{This work was funded by the Brasil 6G Project with support from RNP/MCTI (Grant 01245.010604/2020-14), and by the xGMobile Project (XGM-AFCCT-2025-8-1-1 and XGM-AFCCT-2024-9-1-1) with resources from EMBRAPII/MCTI (Grant 052/2023 PPI IoT/Manufatura 4.0), by CNPq (302085/2025-4, 306199/2025-4, 141545/2024-0), by FAPEMIG (APQ-03162-24, PPE-00124-23, RED-00194-23), and by FINEP (nº 1060/2 contract 01.25.0883.00).
}%
\thanks{M. Khazaee and L. L. Mendes  are with the Radiocommunications
Reference Center, National Institute of Telecommunications (INATEL), Santa Rita do Sapucaí, MG,  37536-001, Brazil (e-mail: \mbox{masoud.khazaee@dtel.inatel.br}; \mbox{lucianol@inatel.br}).
Felipe A. P. de Figueiredo, Rausley A. A. de Souza, and F. D. A. García are with the Wireless and Artificial Intelligence Laboratory (WAI Lab), National Institute of Telecommunications (INATEL), Santa Rita do Sapucaí, MG,  37536-001, Brazil (e-mail: \mbox{felipe.figueiredo@inatel.br}; \mbox{rausley@inatel.br}; \mbox{fernando.almeida@inatel.br}).
F. Rostami Ghadi and K.-K. Wong are with the Department of Electronic and Electrical
Engineering, University College London, Torrington Place, WC1E 7JE, U.K. (e-mail: \mbox{f.rostamighadi@ucl.ac.uk}; \mbox{kai-kit.wong@ucl.ac.uk}).
K.-K. Wong is also with the Department of Electronic Engineering, Kyung Hee University, Yongin-si, Gyeonggi-do 17104, Republic of Korea.
}
}



\maketitle
\begin{abstract}
Fluid reconfigurable intelligent surfaces (FRIS) have recently emerged as a promising technology for enhancing wireless link reliability through spatial decorrelation. 
However, their performance analysis remains challenging due to the sum-product structure of the cascaded channel. 
This letter develops a rigorous analytical framework for FRIS-assisted wireless systems over arbitrarily correlated Nakagami-$m$ fading channels. 
Specifically, we introduce a physically consistent correlation model for \mbox{Nakagami-$m$} fading and derive tractable statistical characterizations for the cascaded channel. 
These results lead to rigorous lower bounds for the outage probability (OP), with a simplified expression also obtained for the independent and identically distributed case. 
To the best of our knowledge, these are the first strict OP lower bounds reported for an FRIS-aided wireless system under arbitrarily correlated Nakagami-$m$ fading. 
CLT- and Gamma-based approximations are included as benchmark methods. Notably, the numerical results show that the proposed OP bound not only provides rigorous performance guarantees but also yields a noticeably tighter OP characterization than the CLT approximation in the high-SNR regime.
\end{abstract}


\section{INTRODUCTION}

\IEEEPARstart{F}{luid} reconfigurable intelligent surfaces (FRIS) have emerged as a promising approach to overcome fundamental limitations of conventional reconfigurable intelligent surface (RIS) architectures. In dense RIS deployments, the close spacing between adjacent elements induces significant spatial correlation, which degrades system performance and limits diversity gains. Moreover, the fixed structure of conventional RIS restricts the exploitation of spatial diversity, even when sufficient physical space is available.
Building on the fluid antenna system (FAS) paradigm~\cite{9264694,10753482,10909643}, fluid reconfigurable intelligent surfaces (FRIS)~\cite{11175437,salem2025lookperformanceenhancementpotential,11075830} extend spatial adaptability by dynamically selecting subsets of weakly correlated elements, thereby improving link reliability.

Despite several recent preliminary works, the channel modeling, optimization, and performance evaluation of RIS-enabled fluid antenna systems and FRIS-assisted communications remain at an early stage of development~\cite{11075830,ghadi2025coverageanalysisoptimizationfiresassisted,11368654,10978677,10539238,11154019Ghadi,Ghadi2025FRISCovert}. 
From a channel modeling and performance analysis perspective, the main challenge lies in deriving the fundamental statistics of the cascaded end-to-end channel gain—particularly its probability density function (PDF) and cumulative distribution function (CDF)—and, consequently, key performance metrics such as outage probability (OP) and average bit error rate (ABER). Due to the sum-product structure of RIS/FRIS cascaded channels and the adopted fading models, obtaining exact or closed-form statistical characterizations is mathematically challenging~\cite{11154019Ghadi,Ghadi2025FRISCovert,GarciaRIS26}.
As a result, most existing works rely on approximation-based approaches or simplified fading assumptions. For example, \cite{10978677} employed a central limit theorem (CLT) and copula-based approximation to characterize the cascaded channel statistics of a RIS-aided fluid antenna system and used the resulting model to evaluate the secrecy outage probability (SOP). Similarly, \cite{10539238} adopted a CLT- and copula-based framework to analyze the OP. More recently, \cite{11154019Ghadi} considered Rayleigh fading and approximated the PDF and CDF of the FRIS cascaded channel using a moment-matched Gamma distribution, enabling the evaluation of OP and ergodic capacity (EC). Building on this framework, \cite{Ghadi2025FRISCovert} investigated FRIS-assisted covert communications under Rayleigh fading, analyzing metrics such as covert outage probability (COP), OP, and success probability (SP). More recently, \cite{khazaee2026exactstatisticalcharacterizationperformance} derived exact expressions for the OP and EC under Rayleigh fading.

Despite their valuable contributions, most existing works rely on asymptotic arguments or distributional approximations. Although such approaches often lead to analytically tractable formulations, they typically do not provide rigorous performance guarantees in the form of strict upper or lower bounds on the system performance. Consequently, the reliability of these approximations can be difficult to assess. Furthermore, most existing studies are restricted to Rayleigh fading, limiting their applicability to more general and realistic propagation environments.
Motivated by these limitations, this paper takes a step toward a more rigorous analytical framework for FRIS-assisted wireless systems. Specifically, we: (i) develop a physically consistent model for correlated Nakagami-$m$ fading channels, which naturally includes Rayleigh fading as a special case when $m=1$; and (ii) derive closed-form lower bounds for the OP, thereby establishing fundamental benchmarks for the achievable system performance.

In the sequel, $\Pr(\cdot)$ and $\mathbb{E}[\cdot]$ denote probability and expectation, respectively; $\|\cdot\|_{2}$ denotes the Euclidean norm; $J_0(\cdot)$ and $K_{\nu}(\cdot)$ denote the zeroth-order Bessel function of the first kind and the modified Bessel function of the second kind of order $\nu$, respectively; $(\cdot)^{\mathrm{T}}$ denotes transpose; ${}_2F_1\!\left(\cdot,\cdot;\cdot;\cdot\right)$ is the Gauss hypergeometric function; $\Gamma(\cdot)$ denotes the Gamma function; and $G_{c,d}^{a,b}(\cdot|\cdot)$ denotes the Meijer $G$-function. Also, $\sim$ means ``distributed as,'' $\mathrm{diag}(\cdot)$ denotes a diagonal matrix, $\mathbf{I}_r$ is the $r\times r$ identity matrix, and $\mathrm{i}\triangleq\sqrt{-1}$. Finally, $\mathcal{G}(a,b)$, $\mathcal{CN}(a,b)$, and $\mathcal{NK}(a,b)$ denote Gamma, circularly symmetric complex Gaussian, and Nakagami-$m$ distributions, respectively.


\section{System Model}
\label{sec: System Model}

\subsection{Surface Geometry and Activation}

We consider a single-antenna base station (BS) communicating with a single-antenna user terminal (UT) assisted by an FRIS comprising $L_{\mathrm{tot}}$ reflecting elements. Due to control constraints, only a subset of these elements is activated for signal reflection.
Let $\mathcal{A}$ denote the set of active elements, with cardinality $|\mathcal{A}| = L_{\mathrm{ON}} \leq L_{\mathrm{tot}}$.
The direct BS–UT link is assumed blocked, such that communication occurs only through the cascaded BS–FRIS and FRIS–UT links.

The FRIS is modeled as a uniform planar array (UPA) with
$M=M_xM_z$ passive elements deployed on the $x$-$z$ plane, where $M_x$
and $M_z$ denote the numbers of elements along the horizontal and vertical
surface directions, respectively. Both axes are measured in meters,
with the origin located at the center of the FRIS surface. The inter-element spacing is given by $d=d_w\lambda_c$, where $\lambda_c$ denotes the carrier wavelength and $d_w$ is the normalized spacing factor.
Due to the close spacing between adjacent elements, spatial correlation must be incorporated into the channel model. Assuming an isotropic Jakes model, the spatial correlation matrix $\boldsymbol{\Sigma}\in\mathbb{C}^{L_{\mathrm{ON}}\times L_{\mathrm{ON}}}$ is given by $[\boldsymbol{\Sigma}]_{r,s}
=
J_0\!\left(
\frac{2\pi d_{r,s}}{\lambda_c}
\right)$, $r,s \in \mathcal{A}$,
where $d_{r,s}=\|\mathbf{r}_r-\mathbf{r}_s\|_2$ is the distance between the $r$-th and $s$-th active FRIS elements.
The active elements are selected using a correlation-aware sub-lattice  element-selection strategy. Let $d_u$ denote the spacing between adjacent candidate elements and $\nu\in\mathbb{N}$ the sub-lattice stride factor. Then, two active elements separated by $p$ horizontal and $q$ vertical sub-lattice steps have distance $d_{p,q}^{(\nu)}=\nu d_u\sqrt{p^2+q^2}$.
To avoid the combinatorial complexity of enforcing decorrelation over all element pairs, we adopt a tractable local-neighbor stencil approach. Specifically, the FRIS placement rule satisfies $\max_{(p,q)\in\mathcal{N}}
\left|
J_0\!\left(
\frac{2\pi \nu d_u}{\lambda_c}\sqrt{p^2+q^2}
\right)
\right|
\le \tau$~\cite[Eq.~(5)]{khazaee2026exactstatisticalcharacterizationperformance},
where $\tau\in(0,1)$ is a prescribed correlation threshold and $\mathcal{N}$ is the stencil that captures the dominant nearest and next-nearest interactions. 

\subsection{Received Signal Model}

Let $\mathbf{h}_1 \triangleq \left[ h_{1,1},\hdots, h_{1,L_{\mathrm{ON}}}\right] \in \mathbb{C}^{L_{\mathrm{ON}}}$ and $\mathbf{h}_2 \triangleq \left[ h_{2,1},\hdots, h_{2,L_{\mathrm{ON}}}\right]\in \mathbb{C}^{L_{\mathrm{ON}}}$ denote the channel vectors associated with the BS--FRIS and FRIS--UT links, respectively. The received baseband signal at the UT is given by
\par\nobreak\vspace{-\abovedisplayskip}
\small
\begin{equation}
y = \sqrt{P}\,\mathbf{h}_2^{H}\boldsymbol{\Phi}\mathbf{h}_1\,x + z,
\label{eq:rx_signal}
\end{equation}
\normalsize
where $x$ is the transmitted symbol with $\mathbb{E}[|x|^2]=1$, $P$ denotes the transmit power, and $z \sim \mathcal{CN}(0,N_0)$ represents additive white Gaussian noise (AWGN) with power $N_0$. The matrix $\boldsymbol{\Phi} = \mathrm{diag}(e^{\mathrm{i} \theta_1},\dots,e^{\mathrm{i} \theta_{L_{\mathrm{ON}}}})$ denotes the FRIS phase-shift matrix.
The channel coefficients are expressed in polar form as $h_{1,\ell} = X_\ell e^{\mathrm{i} \phi_{1,\ell}}$ and $h_{2,\ell} = Y_\ell e^{\mathrm{i}\phi_{2,\ell}}$, for $\ell \in \mathcal{A}$, where $X_\ell$ and $Y_\ell$ denote the fading envelopes, and $\phi_{1,\ell}$ and $\phi_{2,\ell}$ denote the phases of the BS--FRIS and FRIS--UT links, respectively.
We assume Nakagami-$m$ fading, i.e., $X_\ell \sim \mathcal{NK}(m_X,\Omega_{X,\ell})$ and $Y_\ell \sim \mathcal{NK}(m_Y,\Omega_{Y,\ell})$, where $m_X$ and $m_Y$ denote the fading parameters associated with the BS--FRIS and FRIS--UT links, respectively, while $\Omega_{X,\ell}=\mathbb{E}[X_\ell^2]$ and $\Omega_{Y,\ell}=\mathbb{E}[Y_\ell^2]$ denote the corresponding average channel powers, including large-scale attenuation effects for the $\ell$-th BS--FRIS and FRIS--UT links.
Specifically, the average power of the $\ell$-th BS--FRIS link is modeled as
$\Omega_{X,\ell}=\Omega_0(d_0/d_{X,\ell})^{\eta_X}$, where $d_{X,\ell}$ denotes the BS--FRIS distance associated with the $\ell$-th active element, $\Omega_0$ is the reference power at distance $d_0$, and $\eta_X$ is the corresponding path-loss exponent. Similarly, $\Omega_{Y,\ell}=\Omega_0(d_0/d_{Y,\ell})^{\eta_Y}$ models the FRIS--UT link, where $d_{Y,\ell}$ and $\eta_Y$ denote the corresponding distance and path-loss exponent.
The BS--FRIS and FRIS--UT channels are assumed mutually independent, whereas the sequences $\{h_{1,\ell}\}_{\ell=1}^{L_{\mathrm{ON}}}$ and $\{h_{2,\ell}\}_{\ell=1}^{L_{\mathrm{ON}}}$ are spatially correlated due to the finite spacing between FRIS elements; i.e., $h_{k,i}$ and $h_{k,j}$ are correlated for $k \in \{1,2\}$ and $i \neq j$.
The FRIS phase shifts are configured for coherent combining, i.e., $\theta_\ell = -(\phi_{1,\ell}+\phi_{2,\ell})$. Under this phase alignment, the received signal reduces to
\par\nobreak\vspace{-\abovedisplayskip}
\small
\begin{equation}
y = \sqrt{P} S x + z,
\label{eq:aligned_signal}
\end{equation}
\normalsize
where 
\par\nobreak\vspace{-\abovedisplayskip}
\small
\begin{equation}
S \triangleq \sum_{\ell=1}^{L_{\mathrm{ON}}} X_\ell Y_\ell
\label{eq:defS}
\end{equation}
\normalsize
is the effective cascaded channel. Then, the instantaneous signal-to-noise ratio (SNR) at the UT can be expressed as
\par\nobreak\vspace{-\abovedisplayskip}
\small
\begin{equation}
\gamma = \bar{\gamma} S^2,
\label{eq:snr_model}
\end{equation}
\normalsize
where $\bar{\gamma} \triangleq \frac{P}{N_0}$ is the average SNR.

\section{Performance Limits in Nakagami-m Fading}
\label{sec: Performance Limits}

In this section, we first introduce a weighted Cauchy--Schwarz (WCS) framework to derive a tractable upper bound for the effective cascaded channel. Next, we develop a physically consistent correlated Nakagami-$m$ fading model and characterize the resulting random variables through quadratic-form representations. Finally, closed-form lower bounds for the OP are derived.
 
\subsection{Weighted Cauchy--Schwarz Inequality}
\label{sec: CS Inequality}

Define a vector of positive weights $\mathbf{w} \triangleq [w_1,\ldots,w_{L_{\mathrm{ON}}}]^{\mathrm{T}}$. 
Applying the WCS inequality to $S$ in \eqref{eq:defS} yields
\par\nobreak\vspace{-\abovedisplayskip}
\small
\begin{equation}
S^2
\le \mathbf{A}_{\mathbf{w}} \, \mathbf{B}_{\mathbf{w}},
\label{eq:wcs_ineq}
\end{equation}
\normalsize
where $\mathbf{A}_{\mathbf{w}} \triangleq \sum_{\ell=1}^{L_{\mathrm{ON}}} \frac{X_\ell^2}{w_\ell}$ and $\mathbf{B}_{\mathbf{w}} \triangleq \sum_{\ell=1}^{L_{\mathrm{ON}}} w_\ell Y_\ell^2$.
A feasible choice of $\mathbf{w}$ is obtained by minimizing the following deterministic surrogate $\left(\sum_{\ell=1}^{L_{\mathrm{ON}}}\frac{\mathbb{E}[X_\ell^2]}{w_\ell}\right)
\left(\sum_{\ell=1}^{L_{\mathrm{ON}}} \mathbb{E}[Y_\ell^2]\, w_\ell\right)$,
which yields the canonical solution $w_\ell^\star = \sqrt{\mathbb{E}[X_\ell^2]/\mathbb{E}[Y_\ell^2]}$.
Under Nakagami-$m$ fading, this reduces to $w_\ell^\star = \sqrt{\Omega_{X,\ell}/\Omega_{Y,\ell}}$.
After applying this canonical solution into \eqref{eq:wcs_ineq}, we obtain
\par\nobreak\vspace{-\abovedisplayskip}
\small
\begin{equation}
S^2
\le \mathbf{A}_{\mathbf{w}}^\star \mathbf{B}_{\mathbf{w}}^\star \triangleq \mathbf{T}_{\mathbf{w}},
\label{eq:wcs_ineq2}
\end{equation}
\normalsize
where $\textbf{A}_{\textbf{w}}^\star
    =
    \sum_{\ell=1}^{L_{\mathrm{ON}}} \frac{X_\ell^2}{w_\ell^\star}$ and $\textbf{B}_{\textbf{w}}^\star
    =
    \sum_{\ell=1}^{L_{\mathrm{ON}}} w_\ell^\star Y_\ell^2$.

\subsection{Correlated Nakagami-m Fading Model}
\label{sec: Correlation Model}

To facilitate tractable analysis, the Nakagami-$m$ envelopes are represented as sums of squared magnitudes of correlated complex Gaussian random variables. For integer $m_X$, a Nakagami-$m$ random variable $X_\ell$ with parameters $(m_X,\Omega_{X,\ell})$ admits the representation
\par\nobreak\vspace{-\abovedisplayskip}
\small
\begin{equation}
\label{eq: X ell construction}
X_\ell = \sqrt{ \frac{\Omega_{X,\ell}}{m_X} \sum_{k=1}^{m_X} \left| g_{X,k,\ell} \right|^2},
\end{equation}
\normalsize
where $g_{X,k,\ell} \sim \mathcal{CN}(0,1)$ are independent across $k$. For each fixed $k$, the vector $\mathbf{g}_{X,k} \triangleq [g_{X,k,1},\ldots,g_{X,k,L_{\mathrm{ON}}}]^{\mathrm{T}}\sim \mathcal{CN}(\mathbf{0},\boldsymbol{\Sigma})$, thereby modeling the spatial correlation across the FRIS elements. An analogous representation holds for $Y_\ell$ with parameters $(m_Y,\Omega_{Y,\ell})$, where, for each fixed $k$, the vector $\mathbf{g}_{Y,k} \triangleq [g_{Y,k,1},\ldots,g_{Y,k,L_{\mathrm{ON}}}]^{\mathrm{T}} \sim \mathcal{CN}(\mathbf{0},\boldsymbol{\Sigma})$.
Since $\mathbb{E}\!\left[X_\ell\right]=\sqrt{\Omega_\ell/m}\,\Gamma\!\left(m+\frac12\right)/\Gamma(m)$ and the joint moment of two correlated Nakagami-$m$ random variables is available in closed form~\cite{deSouzaYacoub2008BivariateNakagami}, the correlation coefficient between any pair $(X_r,X_s)$ can be obtained as
\par\nobreak\vspace{-\abovedisplayskip}
\small
\begin{align}
\label{eq: corr Nak}
\rho_{r,s}
&=
\frac{\left[\Gamma\!\left(m+\frac12\right)/\Gamma(m)\right]^2 \left[
{}_2F_1\!\left(-\frac12,-\frac12;m;| \left[\boldsymbol{\Sigma}\right]_{r,s}|^2\right)-1
\right]}{m-\left[\Gamma\!\left(m+\frac12\right)/\Gamma(m)\right]^2}.
\end{align}
\normalsize


\textit{Remark 1:} This construction introduces spatial dependence at the level of the underlying complex Gaussian multipath clusters that generate the Nakagami-$m$ envelopes, rather than imposing correlation directly on the resulting envelopes. Consequently, the envelope correlation coefficient in \eqref{eq: corr Nak} emerges from the correlated scattering structure itself. This provides a physically consistent way of relating the Gaussian correlation matrix $\boldsymbol{\Sigma}$ to the induced Nakagami-envelope correlation.



\subsection{Channel Statistics}
\label{sec: Channel Statistics}

Using the Gaussian representation outlined in Section~\ref{sec: Correlation Model},  $\mathbf{A}_{\mathbf{w}}^\star$ and $\mathbf{B}_{\mathbf{w}}^\star$ can be expressed as sums of quadratic forms in correlated Gaussian random vectors, namely, 
\par\nobreak\vspace{-\abovedisplayskip}
\small
\begin{align}
\mathbf{A}_{\mathbf{w}}^\star
&= \sum_{k=1}^{m_X} \mathbf{g}_{X,k}^{H}\mathbf{D}_{\mathbf{A}}\mathbf{g}_{X,k} \label{eq:A_quad} \\
\mathbf{B}_{\mathbf{w}}^\star
&= \sum_{k=1}^{m_Y} \mathbf{g}_{Y,k}^{H}\mathbf{D}_{\mathbf{B}}\mathbf{g}_{Y,k}, \label{eq:B_quad}
\end{align}
\normalsize
where $\mathbf{D}_{\mathbf{A}} 
= \mathrm{diag}\!\left( \frac{\Omega_{X,1}}{m_X w_1^\star}, \ldots, \frac{\Omega_{X,L_{\mathrm{ON}}}}{m_X w_{L_{\mathrm{ON}}}^\star} \right)$ and $\mathbf{D}_{\mathbf{B}} 
= \mathrm{diag}\!\left( \frac{\Omega_{Y,1} w_1^\star}{m_Y}, \ldots, \frac{\Omega_{Y,L_{\mathrm{ON}}} w_{L_{\mathrm{ON}}}^\star}{m_Y} \right)$.
This reformulation is particularly useful because quadratic forms in Gaussian random vectors admit tractable spectral representations. In particular, by applying an eigenvalue decomposition, the quadratic form in \eqref{eq:A_quad} can be expressed as \cite{256521}
\par\nobreak\vspace{-\abovedisplayskip}
\small
\begin{equation}
\mathbf{g}_{X,k}^{H}\mathbf{D}_{\mathbf{A}}\mathbf{g}_{X,k}
=
\sum_{\ell=1}^{L_{\mathrm{ON}}}
\lambda_{\mathbf{A},\ell} |g_{X,k,\ell}|^2,
\label{eq:eigen_quad_A}
\end{equation}
\normalsize
where $\lambda_{\mathbf{A},\ell}$ denotes the $\ell$-th eigenvalue of $\mathbf{D}_{\mathbf{A}}^{1/2}\boldsymbol{\Sigma}\mathbf{D}_{\mathbf{A}}^{1/2}$.

Substituting \eqref{eq:eigen_quad_A} into \eqref{eq:A_quad} and rearranging the order of summation yield $\mathbf{A}_{\mathbf{w}}^\star
=
\sum_{\ell=1}^{L_{\mathrm{ON}}}
U_{\mathbf{A},\ell}$,
where $U_{\mathbf{A},\ell}
\triangleq
\frac{\lambda_{\mathbf{A},\ell}}{m_X}
\sum_{k=1}^{m_X}
|g_{X,k,\ell}|^2$.
Since $|g_{X,k,\ell}|^2 \sim \mathcal{G}(1,1)$ are independent across $k$, the additive property of Gamma random variables implies that $\sum_{k=1}^{m_X} |g_{X,k,\ell}|^2 \sim \mathcal{G}(m_X,1)$. Therefore, by a standard scaling transformation, it follows that $U_{\mathbf{A},\ell}
\sim
\mathcal{G}\!\left(m_X,\frac{\lambda_{\mathbf{A},\ell}}{m_X}\right)$.
Following the same derivation steps used for $\mathbf{A}_{\mathbf{w}}^\star$, $\mathbf{B}_{\mathbf{w}}^\star$ can be expressed as $\mathbf{B}_{\mathbf{w}}^\star
=
\sum_{\ell=1}^{L_{\mathrm{ON}}}
U_{\mathbf{B},\ell}$,
where $U_{\mathbf{B},\ell}
\triangleq
\frac{\lambda_{\mathbf{B},\ell}}{m_Y}
\sum_{k=1}^{m_Y}
|g_{Y,k,\ell}|^2$, and $\lambda_{\mathbf{B},\ell}$ denotes the $\ell$-th eigenvalue of $\mathbf{D}_{\mathbf{B}}^{1/2}\boldsymbol{\Sigma}\mathbf{D}_{\mathbf{B}}^{1/2}$. Since $|g_{Y,k,\ell}|^2 \sim \mathcal{G}(1,1)$ are independent across $k$, it follows that $U_{\mathbf{B},\ell}
\sim
\mathcal{G}\!\left(m_Y,\frac{\lambda_{\mathbf{B},\ell}}{m_Y}\right)$.
Therefore from the distributions of $U_{\mathbf{A},\ell}$ and $U_{\mathbf{B},\ell}$, it follows that $\mathbf{A}_{\mathbf{w}}^\star$ and $\mathbf{B}_{\mathbf{w}}^\star$ are expressed as sums of independent Gamma random variables, with the correlation effects fully captured by the corresponding eigenvalues. 

We now derive the PDFs of $\mathbf{A}_{\mathbf{w}}^\star$ and $\mathbf{B}_{\mathbf{w}}^\star$. Define the parameters $\beta_{\mathbf{A}} \triangleq \frac{1}{m_X}\min_{\ell}\lambda_{\mathbf{A},\ell}$, $\beta_{\mathbf{B}} \triangleq \frac{1}{m_Y}\min_{\ell}\lambda_{\mathbf{B},\ell}$, $\alpha_{\mathbf{A}} \triangleq L_{\mathrm{ON}}m_X$, and $\alpha_{\mathbf{B}} \triangleq L_{\mathrm{ON}}m_Y$. 
By directly applying \cite[eq. (2.9)]{Moschopoulos85}, the PDFs of $\mathbf{A}_{\mathbf{w}}^\star$ and $\mathbf{B}_{\mathbf{w}}^\star$can be written in the unified Gamma-mixture form $f_{\Xi_{\mathbf{w}}^\star}(\xi)
=
\sum_{k=0}^{\infty}
\delta_{\Xi,k}\,
f_G\!\left(\alpha_\Xi+k,\beta_\Xi,\xi\right)$,
where $\Xi \in \{\mathbf{A},\mathbf{B}\}$ and $f_G\!\left(\alpha_\Xi+k,\beta_\Xi,\xi\right)
=
\frac{\xi^{\alpha_\Xi+k-1}}
{\beta_\Xi^{\alpha_\Xi+k}\Gamma(\alpha_\Xi+k)}
\exp\!\left(-\frac{\xi}{\beta_\Xi}\right)$
denotes the PDF of a Gamma random variable with shape parameter $\alpha_\Xi+k$ and scale parameter $\beta_\Xi$.
The coefficients $\delta_{\Xi,k}$ are obtained as
\par\nobreak\vspace{-\abovedisplayskip}
\small
\begin{subequations}
\label{eq:coeff_alpha}
\begin{align}
\delta_{\Xi,0}
&=
\prod_{\ell=1}^{L_{\mathrm{ON}}}
\left(
\frac{m_\varrho\beta_\Xi}{\lambda_{\Xi,\ell}}
\right)^{m_\varrho},
\label{eq:alpha_0}
\\
\delta_{\Xi,k}
&=
\frac{1}{k}
\sum_{i=1}^{k}
\left(
\sum_{\ell=1}^{L_{\mathrm{ON}}}
m_\varrho
\left(
1-\frac{m_\varrho\beta_\Xi}{\lambda_{\Xi,\ell}}
\right)^i
\right)
\delta_{\Xi,k-i},
\quad k\geq1,
\label{eq:alpha_i}
\end{align}
\end{subequations}
\normalsize
where $\varrho \in \{X,Y\}$, with $\varrho=X$ when $\Xi=\mathbf{A}$ and $\varrho=Y$ when $\Xi=\mathbf{B}$.

Since the PDFs of $\mathbf{A}_{\mathbf{w}}^\star$ and $\mathbf{B}_{\mathbf{w}}^\star$ are represented as mixtures of Gamma distributions, the distribution of their product can be derived by directly applying \cite[Proposition~2]{GarciaMixture}. Accordingly, the PDF and CDF of $\mathbf{T}_{\mathbf{w}}$ are given, respectively, by
\par\nobreak\vspace{-\abovedisplayskip}
\small
\begin{align}
\label{eq:fT_expanded}
\nonumber f_{\mathbf{T}_{\mathbf{w}}}(t)
&=
\sum_{k=0}^{\infty}\sum_{j=0}^{\infty}
\frac{2 \, \delta_{\mathbf{A},k}\delta_{\mathbf{B},j} \, \left(\frac{t}{\beta_{\mathbf{A}} \beta_{\mathbf{B}}} \right)^{\frac{1}{2}\left(\alpha_\mathbf{A}+k + \alpha_\mathbf{B}+j\right)}}
{t \, \Gamma(\alpha_\mathbf{A}+k)\Gamma(\alpha_\mathbf{B}+j)} \\
&\quad \times
K_{\alpha_\mathbf{A}+k-\alpha_\mathbf{B}-j}\!\left(2\sqrt{\frac{t}{\beta_{\mathbf{A}} \beta_{\mathbf{B}}}}\right)\\ \label{eq:FT_expanded}
F_{\mathbf{T}_{\mathbf{w}}}(t)
&=
\sum_{k=0}^{\infty}\sum_{j=0}^{\infty}
\delta_{\mathbf{A},k}\delta_{\mathbf{B},j} \mathcal{G}_{k,j}(t),
\end{align}
\normalsize
where
\par\nobreak\vspace{-\abovedisplayskip}
\small
\begin{align}
    \label{eq: Gt}
    \nonumber \mathcal{G}_{k,j}(t) & \triangleq \frac{1}{\Gamma(\alpha_\mathbf{A}+k)\Gamma(\alpha_\mathbf{B}+j)} \\
&\quad \times
G_{1,3}^{2,1}\!\left(
\frac{t}{\beta_{\mathbf{A}} \beta_{\mathbf{B}}}
\;\middle|\;
\begin{matrix}
1\\
\alpha_\mathbf{A}+k,\; \alpha_\mathbf{B}+j,\; 0
\end{matrix}
\right)
\end{align}
\normalsize
denotes the CDF of the product of two independent Gamma random variables with shape parameters $\alpha_\mathbf{A}+k$ and $\alpha_\mathbf{B}+j$, and scale parameters $\beta_\mathbf{A}$ and $\beta_\mathbf{B}$, respectively.

Although \eqref{eq:fT_expanded} and \eqref{eq:FT_expanded} are represented as infinite series, their practical evaluation requires truncation to a finite number of terms, denoted by $\zeta$. 
Thus we introduce the following finite-sum representations of \eqref{eq:fT_expanded} and \eqref{eq:FT_expanded}:
\par\nobreak\vspace{-\abovedisplayskip}
\small
\begin{align}
\bar{f}_{\mathbf{T}_{\mathbf{w}}}(t)
&=
\sum_{k=0}^{\zeta-1}\sum_{j=0}^{\zeta-1}
\frac{2 \, \delta_{\mathbf{A},k} \delta_{\mathbf{B},j}
\left(\frac{t}{\beta_{\mathbf{A}}\beta_{\mathbf{B}}}\right)^{\frac{\alpha_\mathbf{A}+k+\alpha_\mathbf{B}+j}{2}}}
{t\,\Gamma(\alpha_\mathbf{A}+k)\Gamma(\alpha_\mathbf{B}+j)}
\nonumber\\
&\quad\times
K_{\alpha_\mathbf{A}+k-\alpha_\mathbf{B}-j}
\!\left(
2\sqrt{\frac{t}{\beta_{\mathbf{A}}\beta_{\mathbf{B}}}}
\right)
\label{eq:fT_expanded_surrogate}
\\
\bar{F}_{\mathbf{T}_{\mathbf{w}}}(t)
&=
\sum_{k=0}^{\zeta-1}\sum_{j=0}^{\zeta-1}
\delta_{\mathbf{A},k}\delta_{\mathbf{B},j} \mathcal{G}_{k,j}(t).
\label{eq:FT_expanded_surrogate}
\end{align}
\normalsize
Since $\delta_{\mathbf{A},k},\delta_{\mathbf{B},j}\ge 0$ and
$0\le \mathcal{G}_{k,j}(t)\le 1$, the exact CDF can be decomposed as
$F_{\mathbf{T}_{\mathbf{w}}}(t)
=
\bar{F}_{\mathbf{T}_{\mathbf{w}}}(t)
+
R_\zeta(t)$,
where $R_\zeta(t)\ge 0$ represents the residual truncation error associated with the omitted terms. Hence, since $S^2\le \mathbf{T}_{\mathbf{w}}$, the monotonicity of the CDF
implies that the finite-sum representation yields a rigorous finite-truncation
lower bound for any positive integer truncation order $\zeta$, namely,
\par\nobreak\vspace{-\abovedisplayskip}
\small
\begin{align}
    \label{eq: ineq F}
    \bar{F}_{\mathbf{T}_{\mathbf{w}}}(t)\le F_{\mathbf{T}_{\mathbf{w}}}(t)\le F_{S^2}(t),
\end{align}
\normalsize
where $F_{S^2}(\cdot)$ denotes the CDF of $S^2$.
Using $0\le \mathcal{G}_{k,j}(t)\le 1$, the truncation error
$\epsilon_\zeta(t)\triangleq
F_{\mathbf{T}_{\mathbf{w}}}(t)-\bar{F}_{\mathbf{T}_{\mathbf{w}}}(t)$
can be written as
$\epsilon_\zeta(t)
=
\sum_{(k,j)\notin\{0,\ldots,\zeta-1\}^2}
\delta_{\mathbf{A},k}\delta_{\mathbf{B},j}
\mathcal{G}_{k,j}(t)$.
Since all terms in this residual sum are nonnegative, it follows that
$\epsilon_\zeta(t)\ge 0$. Moreover, because each component CDF is upper bounded by one, the truncation error is uniformly bounded by the omitted mixture mass, i.e., $0
\le
\epsilon_\zeta(t)
\le
\sum_{(k,j)\notin\{0,\ldots,\zeta-1\}^2}
\delta_{\mathbf{A},k}\delta_{\mathbf{B},j}$.
The retained mixture mass is
$P_{\mathbf{A}}^{(\zeta)}P_{\mathbf{B}}^{(\zeta)}$, where
$P_{\mathbf{A}}^{(\zeta)}\triangleq
\sum_{k=0}^{\zeta-1}\delta_{\mathbf{A},k}$
and
$P_{\mathbf{B}}^{(\zeta)}\triangleq
\sum_{j=0}^{\zeta-1}\delta_{\mathbf{B},j}$.
Hence, the omitted mixture mass is
$1-P_{\mathbf{A}}^{(\zeta)}P_{\mathbf{B}}^{(\zeta)}$, which yields the truncation bound
\par\nobreak\vspace{-\abovedisplayskip}
\small
\begin{align}
\label{eq: epsilon final}
0
\le
\epsilon_\zeta(t)
\le
1-P_{\mathbf{A}}^{(\zeta)}P_{\mathbf{B}}^{(\zeta)}.
\end{align}
\normalsize
Therefore, the accuracy of the finite-series bound is directly governed by the retained mixture masses
$P_{\mathbf{A}}^{(\zeta)}$ and $P_{\mathbf{B}}^{(\zeta)}$. Thus, \eqref{eq: epsilon final} provides a \emph{sufficient} truncation order to achieve a prescribed accuracy level.

\subsection{Lower Bounds on the Outage Probability}
\label{sec: Fundamental OP Bounds}

The OP is defined as the probability that the received SNR at the UT falls below a prescribed threshold $\gamma_{\mathrm{th}}$, i.e., $P_{\mathrm{out}}
\triangleq
\Pr(\gamma \leq \gamma_{\mathrm{th}})$. By using \eqref{eq:snr_model}, the OP can be rewritten as
\par\nobreak\vspace{-\abovedisplayskip}
\small
\begin{equation}
P_{\mathrm{out}}
=
\Pr\!\left(
S^2 \leq \frac{\gamma_{\mathrm{th}}}{\bar{\gamma}}
\right)= F_{S^2}\!\left(
\frac{\gamma_{\mathrm{th}}}{\bar{\gamma}}
\right),
\label{eq:OP_def}
\end{equation}
\normalsize
Recalling that $S^2 \leq \mathbf{T}_{\mathbf{w}}$ and combining this relation with \eqref{eq: ineq F}, the monotonicity of the CDF implies that
\par\nobreak\vspace{-\abovedisplayskip}
\small
\begin{align}
    \label{eq:OP ineq}
    \bar{F}_{\mathbf{T}_{\mathbf{w}}}\!\left(
\frac{\gamma_{\mathrm{th}}}{\bar{\gamma}}
\right) \leq F_{\mathbf{T}_{\mathbf{w}}}\!\left(
\frac{\gamma_{\mathrm{th}}}{\bar{\gamma}}
\right)
\leq
F_{S^2}\!\left(
\frac{\gamma_{\mathrm{th}}}{\bar{\gamma}}
\right).
\end{align}
\normalsize

Therefore, substituting \eqref{eq:FT_expanded_surrogate} into
\eqref{eq:OP ineq} yields the following rigorous finite-truncation
closed-form lower bound for the OP:
\par\nobreak\vspace{-\abovedisplayskip}
\small
\begin{align}
P_{\mathrm{out}}
&\geq
\sum_{k=0}^{\zeta-1}\sum_{j=0}^{\zeta-1} \delta_{\mathbf{A},k}\delta_{\mathbf{B},j} \mathcal{G}_{k,j} \left(\frac{\gamma_{\mathrm{th}}}{\bar{\gamma}} \right).
\label{eq:final_OP}
\end{align}
\normalsize

If the BS--FRIS and FRIS--UT links are independent and identically distributed (i.i.d.) across the FRIS elements, then $\Omega_{X,\ell}=\Omega_X$, $\Omega_{Y,\ell}=\Omega_Y$, and $w_\ell^\star=\sqrt{\Omega_X/\Omega_Y}$ for all $\ell$. In addition, for independent elements, $\boldsymbol{\Sigma}=\mathbf{I}_{L_{\mathrm{ON}}}$, so that all eigenvalues are identical, yielding $\lambda_{\mathbf{A},\ell}=\sqrt{\Omega_X\Omega_Y}/m_X$ and $\lambda_{\mathbf{B},\ell}=\sqrt{\Omega_X\Omega_Y}/m_Y$ for all $\ell$. Consequently, the  coefficients reduce to $\delta_{\Xi,0}=1$ and $\delta_{\Xi,k}=0$ for $k\geq1$. Therefore, under i.i.d. Nakagami-$m$ fading, \eqref{eq:final_OP} simplifies to the following closed-form OP lower bound:
\par\nobreak\vspace{-\abovedisplayskip}
\small
\begin{align}
P_{\mathrm{out}}
&\geq
\frac{1}
{\Gamma(\alpha_\mathbf{A})\Gamma(\alpha_\mathbf{B})}
G_{1,3}^{2,1}\!\left(
\frac{\gamma_{\mathrm{th}} m_X m_Y}
{\bar{\gamma}\Omega_X\Omega_Y}
\;\middle|\;
\begin{matrix}
1\\
\alpha_\mathbf{A},\; \alpha_\mathbf{B},\; 0
\end{matrix}
\right).
\label{eq:final_OP_iid}
\end{align}
\normalsize

\textit{Remark 2:} To the best of our knowledge, \eqref{eq:final_OP} and \eqref{eq:final_OP_iid} represent the first strict lower bounds reported for the OP of an FRIS-aided wireless system under arbitrarily correlated and i.i.d. \mbox{Nakagami-$m$} fading, respectively. These results establish fundamental benchmarks for the minimum attainable OP under the adopted channel models. Unlike existing CLT- and moment-matched-based FRIS analyses, which rely on asymptotic conditions or heuristic approximations, the proposed framework provides rigorous outage bounds with provable guarantees for arbitrary numbers of FRIS elements, without imposing predefined distributional structures on the cascaded channel. Notably, \eqref{eq:final_OP} provides substantially more accurate OP estimates than the CLT approximation in the high-SNR regime, as confirmed in Section~\ref{sec:numerical_results}.

Note that the derived PDF, CDF, and OP expressions remain valid for general Nakagami-$m$ fading with arbitrary positive real parameters $m_X,m_Y>0$, since the analysis ultimately relies on the Moschopoulos Gamma-sum representation~\cite{Moschopoulos85}, which holds for arbitrary positive real shape parameters.

\section{Numerical Results}
\label{sec:numerical_results}

\begin{figure}[!t]
\centering
\includegraphics[width=1.0\linewidth]{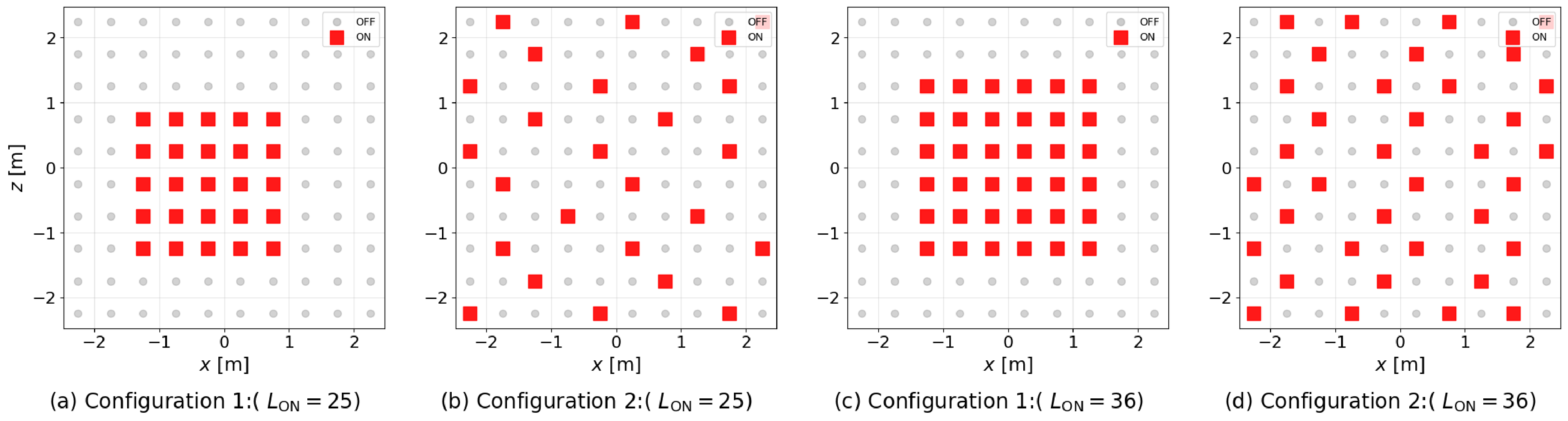}
\caption{Element activation patterns.}
\label{fig:element_activation}
\end{figure}

This section validates the analytical OP expressions through Monte Carlo (MC) simulations for a $10\times10$ FRIS surface with $L_{\mathrm{ON}}\in\{25,36\}$. CLT- and Gamma-based approximations are included as benchmark methods. For each activation size, two active-element configurations are considered: Configuration~1 adopts a compact central selection resembling a conventional RIS, whereas Configuration~2 employs the proposed correlation-aware selection with $\mathcal{N}=\{(1,0),(1,1),(2,0),(2,1),(2,2),(3,0)\}$ and $\tau=0.4$, resulting in the 2-dimensional (2-D) geometry shown in Fig.~\ref{fig:element_activation}. 
Moreover, we set $\Omega_0=d_0=1$, $\eta_X=\eta_Y=2.2$, and, for MC simulation purposes, $m_X=2$ and $m_Y=3$.
To ensure $\epsilon_\zeta(t) \leq 1-P_{\mathbf{A}}^{(\zeta)}P_{\mathbf{B}}^{(\zeta)} \leq 10^{-3}$ across all considered scenarios, we set the truncation order to $\zeta=5000$. This choice guarantees the prescribed accuracy and should be regarded as a sufficient, rather than necessary, truncation order. In practice, fewer terms may be sufficient.
In the adopted 2-D geometry, the BS and UT are located at $(x,z)=(0,-5)$~m and $(x,z)=(0,5)$~m, respectively, where the origin is placed at the center of the FRIS surface.

\begin{figure}[!t]
\centering
\includegraphics[width=0.82\linewidth]{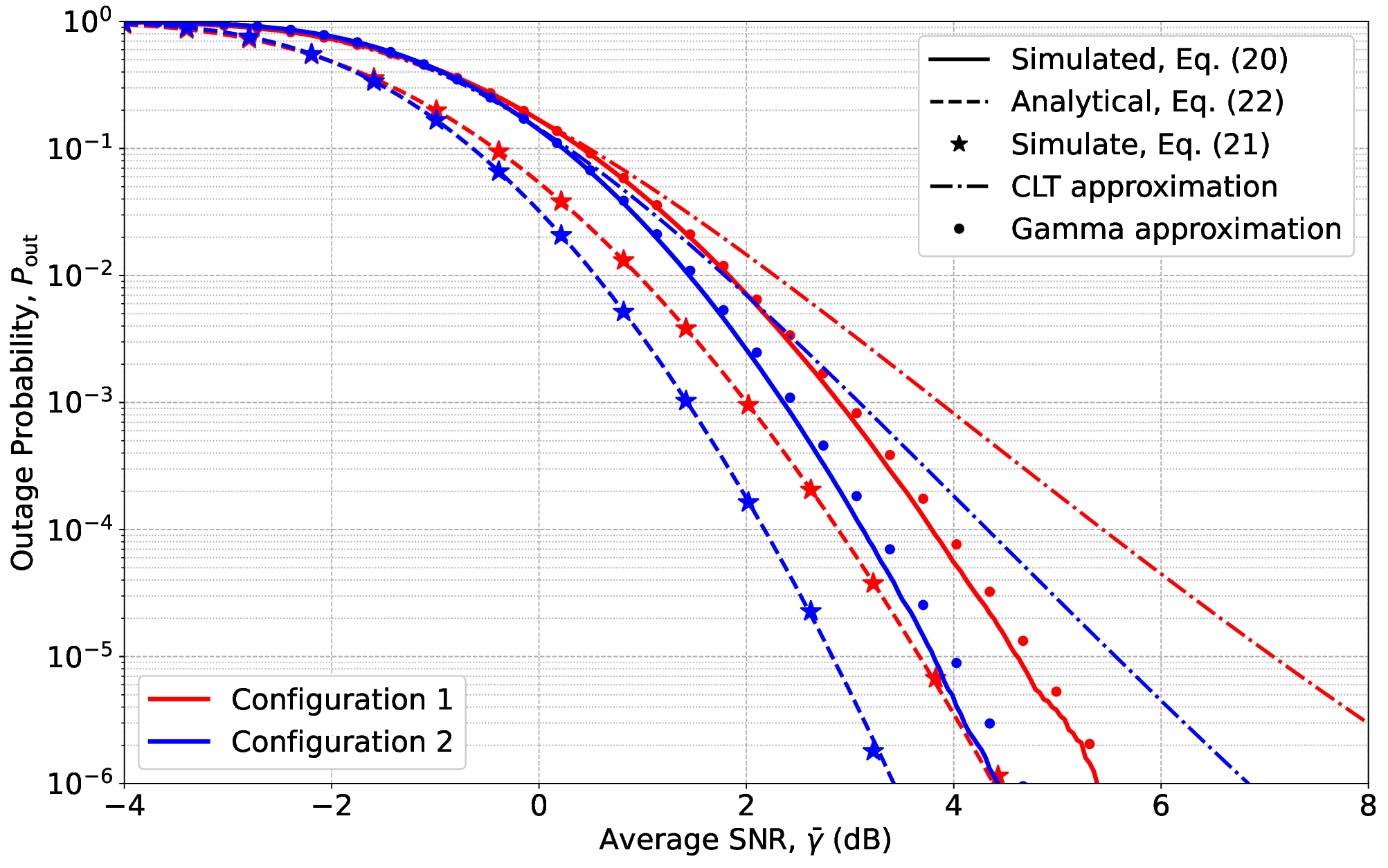}
\caption{OP versus the average SNR $\bar{\gamma}$ for $L_{\mathrm{ON}}=25$.}
\label{fig:op_l25}
\end{figure}

\begin{figure}[!t]
\centering
\includegraphics[width=0.82\linewidth]{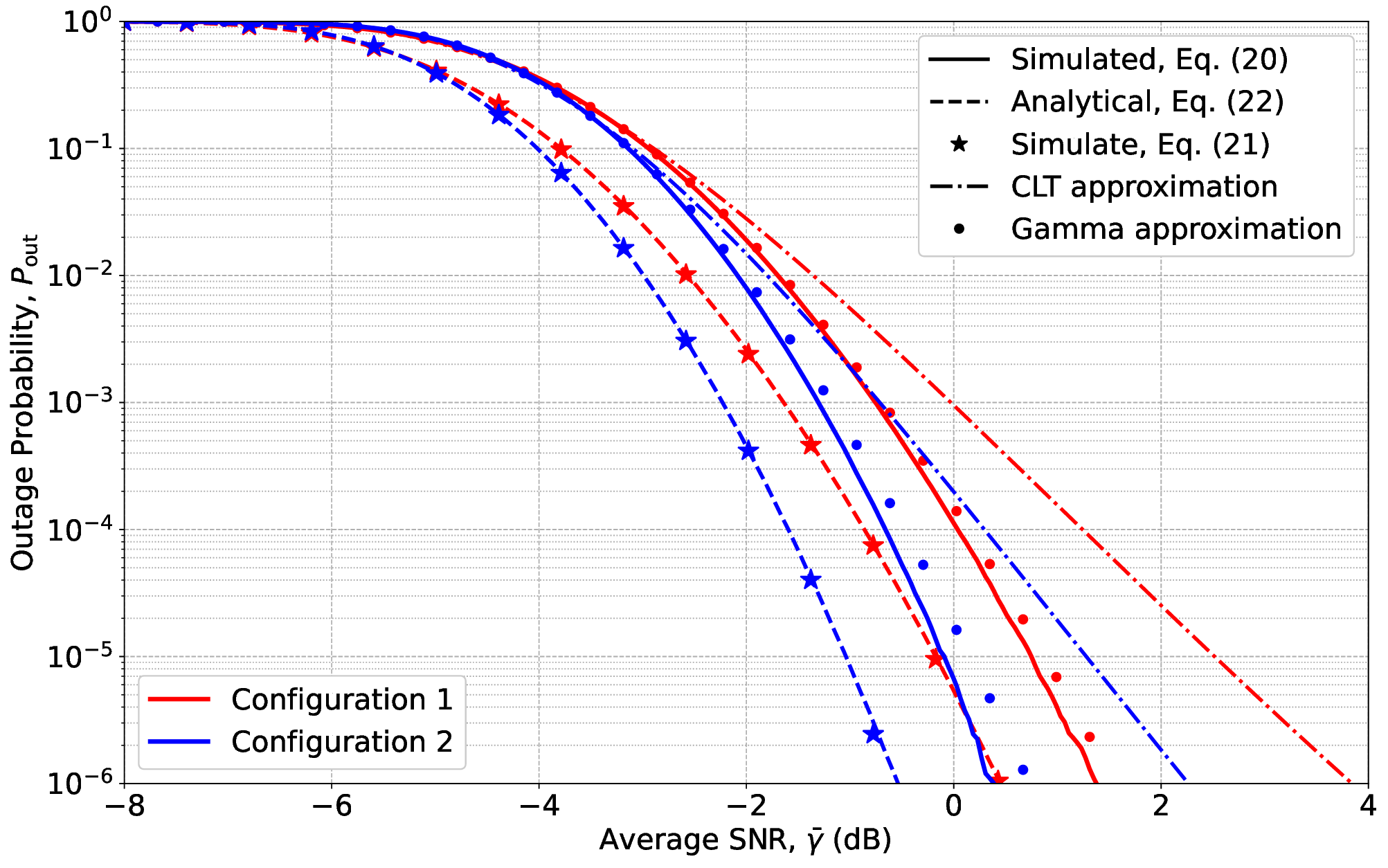}
\caption{OP versus the average SNR $\bar{\gamma}$ for $L_{\mathrm{ON}}=36$.}
\label{OP_vs_SNR36}
\end{figure}

Figs.~\ref{fig:op_l25} and~\ref{OP_vs_SNR36} present the OP results for $L_{\mathrm{ON}}=25$ and $L_{\mathrm{ON}}=36$, respectively. In both cases, the analytical curves from \eqref{eq:final_OP} closely match the MC simulations, confirming the accuracy of the proposed derivation. Configuration~2 outperforms Configuration~1 because the correlation-aware FRIS activation selects weakly correlated elements, thereby improving spatial diversity. In addition, increasing $L_{\mathrm{ON}}$ from $25$ to $36$ shifts the OP curves toward lower average SNR values, demonstrating the performance gain achieved by activating more FRIS elements.
Figs.~\ref{fig:op_l25} and~\ref{OP_vs_SNR36} further show that the CLT and Gamma approximations do not provide rigorous performance guarantees. Although these approximations may agree with the MC results over limited SNR ranges, they can appear either above or below the true OP curves, and therefore cannot be interpreted as valid upper or lower bounds. In contrast, the proposed bound remains consistently below the MC results, as theoretically expected, while closely tracking the true OP behavior. Notably, it provides a tighter OP characterization than the CLT approximation in the high-SNR region.

\section{Conclusion}
\label{sec: Conclusion}
This letter developed an analytical framework for FRIS-assisted wireless systems over arbitrarily correlated Nakagami-$m$ fading channels. 
By combining a WCS framework with a physically consistent correlation model, tractable cascaded-channel statistics were obtained. 
Based on these results, rigorous lower bounds for the OP were derived, avoiding heuristic and asymptotic distributional approximations.
Numerical results confirmed the validity of the proposed bounds, with particularly tight performance at low SNR.
Overall, the proposed framework establishes fundamental outage-performance benchmarks beyond Rayleigh fading.

\bibliographystyle{IEEEtran}
\bibliography{references} 

@article{deSouzaYacoub2008BivariateNakagami,
  author  = {{de Souza}, Rausley A. A. and Yacoub, Michel D.},
  title   = {Bivariate {Nakagami-$m$} Distribution With Arbitrary Correlation and Fading Parameters},
  journal = {IEEE Trans. Wireless Commun.},
  volume  = {7},
  number  = {12},
  pages   = {5227--5232},
  month   = {Dec.},
  year    = {2008},
  doi     = {10.1109/T-WC.2008.071152}
}

@ARTICLE{GarciaRIS26,
  author={García, Fernando Darío Almeida and Calmon, Flavio du Pin and Filho, José Cândido Silveira Santos},
  journal={IEEE Trans. Veh. Technol.}, 
  title={Enhanced Channel Characterization for {RIS}-Aided Systems Over {N}akagami-{$m$} Fading Channels}, 
  year={2026},
  month={Apr.},
  volume={75},
  number={4},
  pages={6207-6223},
  doi={10.1109/TVT.2025.3622360}}

@ARTICLE{GarciaMixture,
  author={García, Fernando Darío Almeida and Daoud Yacoub, Michel and de Figueiredo, Felipe A. P. and Adriano Amaral de Souza, Rausley},
  journal={IEEE Trans. Commun.}, 
  title={ $\alpha$-$\mathcal{F}$  {M}ixture and  $\alpha$-$\mathcal{G}$  {M}ixture: Unifying Composite Fading Models}, 
  year={2026},
  month={Mar.},
  volume={74},
  number={},
  pages={6059-6074},
  doi={10.1109/TCOMM.2026.3672165}}

@inproceedings{Ghadi2025FRISCovert,
  author    = {F. R. Ghadi and M. Kaveh and H. Hong and K.-K. Wong and R. J{\"a}ntti and F. J. L{\'o}pez-Mart{\'\i}nez},
  title     = {Performance Analysis of Fluid Reconfigurable Intelligent Surface Over Covert Communications},
  booktitle = {Proc. 4th Int. Conf. {6G} Netw. (6GNet)},
  month={Dec.},
  address   = {Paris, France},
  year      = {2025},
  pages     = {82-86},
  doi       = {10.1109/6GNet68413.2025.11314078}
}

@article{Moschopoulos85, 
	Author = {P. G. Moschopoulos},
	Journal = {Ann. Inst. Statist. Math. (Part A)},
	Number = {},
	Pages = {541–544},
	Title = {The distribution of the sum of independent {G}amma random variables},
	Volume = {37},
	Month = {},
	Year = {1985}}

@ARTICLE{11154019Ghadi,
  author={Ghadi, Farshad Rostami and Wong, Kai-Kit and López-Martínez, F. Javier and Alexandropoulos, George C. and Chae, Chan-Byoung},
  journal={IEEE Wireless Commun. Lett.}, 
  title={Performance Analysis of Wireless Communication Systems Assisted by Fluid Reconfigurable Intelligent Surfaces}, 
  year={2025},
  month={Sep.},
  volume={},
  number={},
  pages={1-1},
  doi={10.1109/LWC.2025.3608040}}

@ARTICLE{9264694,
  author={Wong, Kai-Kit and Shojaeifard, Arman and Tong, Kin-Fai and Zhang, Yangyang},
  journal={ IEEE Trans. Wireless Commun.}, 
  title={Fluid Antenna Systems}, 
  year={2021},
  month={Nov.},
  volume={20},
  number={3},
  pages={1950-1962},
  doi={10.1109/TWC.2020.3037595}}

@ARTICLE{11075830,
  author={Xiao, Han and Hu, Xiaoyan and Wong, Kai-Kit and Hong, Hanjiang and Alexandropoulos, George C. and Chae, Chan-Byoung},
  journal={IEEE Wireless Commun. Lett.}, 
  title={Fluid Reconfigurable Intelligent Surfaces: Joint On-Off Selection and Beamforming With Discrete Phase Shifts}, 
  year={2025},
  month={Jul.},
  volume={14},
  number={10},
  pages={3124-3128},
  doi={10.1109/LWC.2025.3587070}}

@INPROCEEDINGS{10978677,
  author={Ghadi, Farshad Rostami and Wong, Kai-Kit and Kaveh, Masoud and Lopez-Martinez, F. Javier and New, Wee Kiat and Xu, Hao},
  booktitle = {IEEE Wireless. Commun. Netw. Conf. WCNC},
  title={Secrecy Performance Analysis of {RIS}-Aided Fluid Antenna Systems}, 
  year={2025},
  month={May},
  volume={},
  number={},
  pages={1-6},
  doi={10.1109/WCNC61545.2025.10978677}}

@ARTICLE{10539238,
  author={Rostami Ghadi, Farshad and Wong, Kai-Kit and New, Wee Kiat and Xu, Hao and Murch, Ross and Zhang, Yangyang},
  journal={IEEE Wireless Commun. Lett.}, 
  title={On Performance of {RIS}-Aided Fluid Antenna Systems}, 
  year={2024},
  month={May},
  volume={13},
  number={8},
  pages={2175-2179},
  doi={10.1109/LWC.2024.3405636}}

@misc{salem2025lookperformanceenhancementpotential,
      title={A First Look at the Performance Enhancement Potential of Fluid Reconfigurable Intelligent Surface}, 
      author={Abdelhamid Salem and Kai-Kit Wong and George Alexandropoulos and Chan-Byoung Chae and Ross Murch},
      year={2025},
      eprint={2502.17116},
      archivePrefix={arXiv},
      primaryClass={eess.SP},
      url={https://arxiv.org/abs/2502.17116}, 
}

@ARTICLE{11368654,
  author={Kaveh, Masoud and Rostami Ghadi, Farshad and Hernando-Gallego, Francisco and Martín, Diego and Wong, Kai-Kit and Jäntti, Riku},
  journal={IEEE Wirel. Commun.}, 
  title={Physical Layer Security Over Fluid Reconfigurable Intelligent Surface-Assisted Communication Systems}, 
  year={2026},
  month={Jan},
  volume={15},
  number={},
  pages={1697-1701},
  doi={10.1109/LWC.2026.3659514}}

@misc{ghadi2025coverageanalysisoptimizationfiresassisted,
      title={Coverage Analysis and Optimization of FIRES-Assisted {NOMA} and {OMA} Systems}, 
      author={Farshad Rostami Ghadi and Kai-Kit Wong and Masoud Kaveh and Hanjiang Hong and Chan-Byoung Chae and Lajos Hanzo},
      year={2025},
      eprint={2511.01111},
      archivePrefix={arXiv},
      primaryClass={cs.IT},
      url={https://arxiv.org/abs/2511.01111}, 
}

@ARTICLE{10753482,
  author={New, Wee Kiat and Wong, Kai-Kit and Xu, Hao and Wang, Chao and Ghadi, Farshad Rostami and Zhang, Jichen and Rao, Junhui and Murch, Ross and Ramírez-Espinosa, Pablo and Morales-Jimenez, David and Chae, Chan-Byoung and Tong, Kin-Fai},
  journal={IEEE Commun. Surv. Tutor}, 
  title={A Tutorial on Fluid Antenna System for {6G} Networks: Encompassing Communication Theory, Optimization Methods and Hardware Designs}, 
  year={2025},
  month={Nov.},
  volume={27},
  number={4},
  pages={2325-2377},
  doi={10.1109/COMST.2024.3498855}}

@ARTICLE{10909643,
  author={Lu, Wen-Jun and He, Chun-Xing and Zhu, Yongxu and Tong, Kin-Fai and Wong, Kai-Kit and Shin, Hyundong and Cui, Tie Jun},
  journal={IEEE Commun Mag.}, 
  title={Fluid Antennas: Reshaping Intrinsic Properties for Flexible Radiation Characteristics in Intelligent Wireless Networks}, 
  year={2025},
  month={Mar.},
  volume={63},
  number={5},
  pages={40-45},
  doi={10.1109/MCOM.002.2400490}}

@ARTICLE{11175437,
  author={Hong, Hanjiang and Wong, Kai-Kit and Chae, Chan-Byoung and Xu, Hao and Guo, Xinghao and Ghadi, Farshad Rostami and Chen, Yu and Xu, Yin and Liu, Baiyang and Tong, Kin-Fai and Zhang, Yangyang},
  journal={IEEE Trans. Netw. Sci. Eng.}, 
  title={A Contemporary Survey on Fluid Antenna Systems: Fundamentals and Networking Perspectives}, 
  year={2026},
  month={Sep},
  volume={13},
  number={},
  pages={2305-2328},
  doi={10.1109/TNSE.2025.3613225}}

@misc{khazaee2026exactstatisticalcharacterizationperformance,
      title={Exact Statistical Characterization and Performance Analysis of Fluid Reconfigurable Intelligent Surfaces}, 
      author={Masoud Khazaee et al.},
      year={2026},
      eprint={2603.28974},
      archivePrefix={arXiv},
      primaryClass={eess.SP},
      url={https://arxiv.org/abs/2603.28974}, 
}

@ARTICLE{256521,
  author={Biyari, K.H. and Lindsey, W.C.},
  journal={IEEE Transactions on Information Theory}, 
  title={Statistical distributions of Hermitian quadratic forms in complex Gaussian variables}, 
  year={1993},
  volume={39},
  number={3},
  pages={1076-1082},
  doi={10.1109/18.256521}}

\end{document}